\documentclass[sageh,Royal,times]{sagej}

\usepackage{url}
\usepackage{eurosym}
\usepackage{balance}
\usepackage{natbib}
\usepackage{tabularx}
\bibpunct[, ]{(}{)}{,}{a}{}{,}

\usepackage[colorlinks=true, linkcolor=black,citecolor=black,urlcolor=black]{hyperref}

\newcommand{\change}[1]{\textcolor{black}{#1}}

\begin{document}

\title{David and Goliath: Privacy Lobbying in the European Union}

\author{Jukka Ruohonen\affilnum{1}}
\affiliation{\affilnum{1} Department of Future Technologies, University of Turku, FI-20014 Turun yliopisto, Finland}
\corrauth{Jukka Ruohonen}
\email{juanruo@utu.fi}


\begin{abstract}
The paper examines a question of how much more resources do organized business interests have when compared to resources of civil society groups in the context of privacy lobbying in the European Union (EU). To answer to the question, the paper draws from classical literature on power resources and pluralism. The empirical material comes from a lobbying register maintained by the EU. According to the results, (a)~there is only a small difference in terms of the average financial and human resources, but a vast difference when absolute amounts are used. Furthermore, (b)~organized business interests are better affiliated with each other and other organizations. Finally, (c) many organized business interests maintain their offices in the United States, whereas the non-governmental organizations observed are mostly European. With these results and the accompanying discussion, the paper contributes to the underresearched  but inflammatory topic of privacy politics.
\end{abstract}


\maketitle

\section*{Introduction}

The ever so famous General Data Protection Regulation (GDPR) was enforced in the European Union in May 2018. Even though vast amounts have been written about this particular regulation, the political aspects have received surprisingly little attention. The point applies also to scholarly research. It is trivial to find thousands of recent research papers, position papers, conference presentations, technical reports, commentaries, and related outputs from scholarly work across sciences and humanities. However, social and political scientists have remained mostly \text{silent---to} cast aside the commentaries about the regulation's perceived impact upon their own craft. This silence is unexpected because the GDPR is also about power---about the fundamental concept in political science.

Resources to exercise power are the subject of this paper. The paper continues the recent work of \citet{Minkkinen19} on approaching privacy policies through privacy politics and political institutions. The European Union is the polity of interest. While Minkkinen's focus was on institutions through which narratives for the future are being told, the present work takes a much more direct, mundane, and even slightly cynical standpoint on privacy politics. The research question examined is simple and free from any evasiveness: how much more power resources do industry representatives have compared to civil society groups when privacy issues are lobbied in the European Union?

In essence, power resources are about the bases of power rather than about the exercise of power, and ``from basic power resources actors can derive other power resources'' in order to ensure ``the capacity to reward or to punish other actors''~\citep[p.~34]{Korpi85}. Here, the capacity to punish or reward refers to the exercising of power through resources that enable it. The power itself is understood in relational terms; according to a classical definition, A has power over B when A can influence B to do something that B would not otherwise do~\citep{Dahl57}. Adjectives follow. ``By powerful we mean, of course, those who are able to realize their will, even if others resist it'' \citep[p.~9]{Mills56}. Stated this way, the concept may seem deceivingly simple, which it is not, of course. The fundamental problems relate to the relational characteristics, which are difficult to conceptualize and measure~\citep{Dahl57, Dur08}. At the expense of depth and realism, the focus on power resources avoids these major problems. Given many transparency improvements, measuring lobbying (power) resources in the EU have greatly improved.

Privacy is easy to justify as a highly relevant topic for lobbying research. With few exceptions such environmental politics, nowhere has lobbying been so prevalent in recent EU politics. The European lobbying boom started in the late 1980s and early 1990s during the creation of the Single Market and the gradual transfer of the regulatory power of the member states to the EU-level~\citep{Broscheid03}. Lobbying later evolved into a well-established political practice for policy making in the union, often exercised between full-time lobbyists and politicians in rooms with more or less closed doors. These political rooms include also various high-level meetings and related arrangements. The 2010s privacy politics are among the prime examples on how the closed doors were at least partially opened; on how the lobbying practices became known also among the general public. Particularly social media was on flames during the policy making of the GDPR, and the fire has spread to the current conflagration of privacy politics. Whether it is non-governmental organizations~\citep{EDRI19a}, public data protection authorities, or academics, the fierce lobbying by organized business interests has left everyone dazzled. ``Their alarmism at times loses all proportion'' \citep[p.~275]{Dix13}. At the same time, many technology giants have started to exploit a privacy rhetoric to legitimize their interests~\citep{Lindh17}. These political discourses surrounding privacy lobbying are interesting---yet no previous work has been done to examine the resources required to upkeep the discourses. The same applies to closed-door lobbying.

\section*{Backstage}

\subsection{Pluralism}

Privacy and data protection are fundamental questions of the 21st century. For many observers, the GDPR most of all ``represents the ongoing battle between unfettered capitalism and human dignity'' \citep{Greengard18}. The keywords of battle, capitalism, and dignity might spell out from a random monograph addressing the great narratives of the 20th century politics. But as it stands, these are also the underlying keywords in \citeauthor{Zuboff19}'s recent \citeyearpar{Zuboff19} monumental monograph; a grand theory on the evolution of capitalism toward surveillance coupled with a rather pessimistic stance on the possibilities to resist this evolutionary trajectory. While \citeauthor{Zuboff19}'s work certainly deserves a thorough assessment also in political science, it suffices here to underline that the work more or less continues the tradition of so-called surveillance studies exercised particularly in sociology and related disciplines. At the risk of an overstatement, it can be remarked that this research tradition and its variants are characteristically theoretical and occasionally even hostile toward empirical research~\citep[cf.][]{Bohme19, Clarke19, Cohen15, Marx07}. The present work contributes to the attempts to seal this gap in the existing literature. But when empirical research is a goal, how theoretical framing might be done with respect to the equally difficult concepts of privacy and power?

A traditional comparative path would lead to consider the origin of the GDPR, the European Union, in relation to other powers, including the United States in particular. This comparative path seems sensible at first glance. After all, ``the United States versus the world'' has long been the global setup in both privacy politics and the resulting privacy regulations~\citep[cf.][]{Geller16}. Most large companies at the center of recent privacy controversies originate from the United States. This has created an enduring schism in the EU, which, however, does not explain the rationale behind the GDPR. Rather, the rationale traces to the different underpinnings of many fundamental concepts in the legal and philosophical traditions characterizing the United States and Europe. Human dignity is one of these concepts~\citep{Neier13}. Self-determination is another~\citep{Bohme19, Rouvroy09}. The long-standing controversies between the United States and the European Union regarding the historical Safe Harbor Principles and the present Privacy Shield framework would further justify the comparative path. By continuing further on the path, also many famous theoretical frameworks unveil themselves. 

In particular, the comparative path would sooner or later lead to the so-called varieties-of-capitalism (VoC) school of thought and its framing between liberal (uncoordinated) and coordinated market economies~\citep{Korpi06, MartinThelen07, Soskice99}. As temping as this path is, it contains several severe limitations in the privacy context. 

In terms of theory, the VoC approach has always emphasized corporatism as the most important frame for interest groups and collective action. Corporatism has little to do with privacy, but the problem runs deeper. Many themes in the theory of corporatism relate to industrial citizenship---a combination of civil, political, and social rights---and its \textit{collective} enforcement through the freedom of contract~\citep{Streeck97}. While industrial citizenship is part of human rights, privacy belongs to the so-called third wave of fundamental rights that go beyond the traditional civil, political, and social rights. Nor is it possible to collectively bargain over someone's privacy through contractual exchange. In terms of empirical research, the VoC approach is problematic due to the lack of robust comparative data on privacy, whether in terms of regulations or violations. In terms of practice, it is unclear whether a distinction between uncoordinated and coordinated approaches to privacy is sharp enough for gauging the present reality. While it remains to be seen whether the dominant form in the United States---industry self-regulation---might be still salvaged~\citep{Listokin17}, it should be understood that privacy has had a legislative grounding also in the United States; the Privacy Act of 1974 is the prime example in this regard. At the time of writing, it seems also likely that a federal privacy law will be enacted in the future partially due to the GDPR's much debated extraterritorial nature~\citep{Isaak18}. Despite of these and other problems, the VoC approach is not without its merits also in the context of privacy: it forces to think in terms of interest groups and collective action. Another merit is the approach's long history in specifically emphasizing the role and power of employer associations and organized business interests in general~\citep{Korpi06, MartinSwank12}. As will be elaborated, business interests are an essential part of privacy politics, although the political construction of these has fundamentally transformed from the traditional corporatist setup for coordination. Therefore, in order to think in these VoC terms, the theory of corporatism must be switched to its classical rival, the theory of pluralism.

Privacy politics are not exercised solely and primarily \textit{between} political parties. Instead, Zuboff's ``fight for a human future at the new frontier of power'' occurs through a complex constellation of civil society groups, technology giants~\citep{Popiel18}, industry associations and business groups, different intermediaries \citep{Zajko18}, such as Internet service providers and content delivery networks, standardization bodies and funding agencies~\citep{Mueller19}, and other stakeholders, including also individuals, whether engineers, scholars, consumers, politicians, bureaucrats, or lobbyists. This constellation is almost like a textbook definition for pluralism: power is scattered across multiple distinct agencies, but policy outcomes (privacy regulations) are delivered and executed \textit{through} political parties and classical democratic institutions. However, the scattering of power is not evenly distributed. One agent may hold power, but another agent may hold more power. A less powerful agent may also have resources to become more powerful, and yet even a powerful agent may be poor at exercising its power. These basic characterizations of the pluralist power theory are important for better understanding the lobbying for or against privacy in the European Union and elsewhere.

\subsection{Lobbying}

The impact of interest groups on policy outcomes is a classical topic in political science. The scholarly history includes many modern classics, including \citeauthor{Dahl61}'s \citeyearpar{Dahl61} seminal work on pluralism. According to his classical pluralists approach, many interest groups possess at least some power to exert influence over policy outcomes even in purely parliamentary political systems. The sources of power vary for influencing others in political systems. The examples include, but are not limited to, money, legitimacy, political backing, age, gender, ethnicity, knowledge, expertise, prestige, and charisma.

These and other sources of power are used to influence policy making in various stages of the legislative process in the EU. The structure of a political system shapes the organization of interest groups~\citep{Naoi09}. Different EU institutions are also approached and lobbied differently: the different roles of the Commission, the Parliament, and the Council require different lobbying strategies \citep{Dur09, Massaro19}. The Parliament is the only EU institution that is directly elected and thus accountable to the EU electorate; the members of the Commission and Council are appointed. Whether democratic accountability actually occurs in practice is a subject of a long-standing debate. In addition to the extensive literature on the EU's general democratic deficit, it may be that the EU electorate mostly holds national parties and their leaderships accountable instead of the Members of the European Parliament~\citep{Mahoney07}. National parties remain also important targets for lobbying by interest groups. Although the influence of national parties on EU policies has decreased, many interest groups continue to lobby also national parties even on EU matters~\citep{Rasmussen12}. With regards to the Parliament itself, previous results further indicate that lobbying tends to follow traditional party lines, ideologies, and power relations~\citep{DeBruycker16}. To some extent, these points support also the pluralist theory: there continues to be a parliamentary democratic deficit, but interests are still channeled to the supranational level despite of weak EU parties and divided institutions~\citep{Coultrap99}. There are five important further points about this pluralist channeling of political interests in the EU.

\begin{figure}[th!b]
\centering
\includegraphics[width=12cm, height=4.5cm]{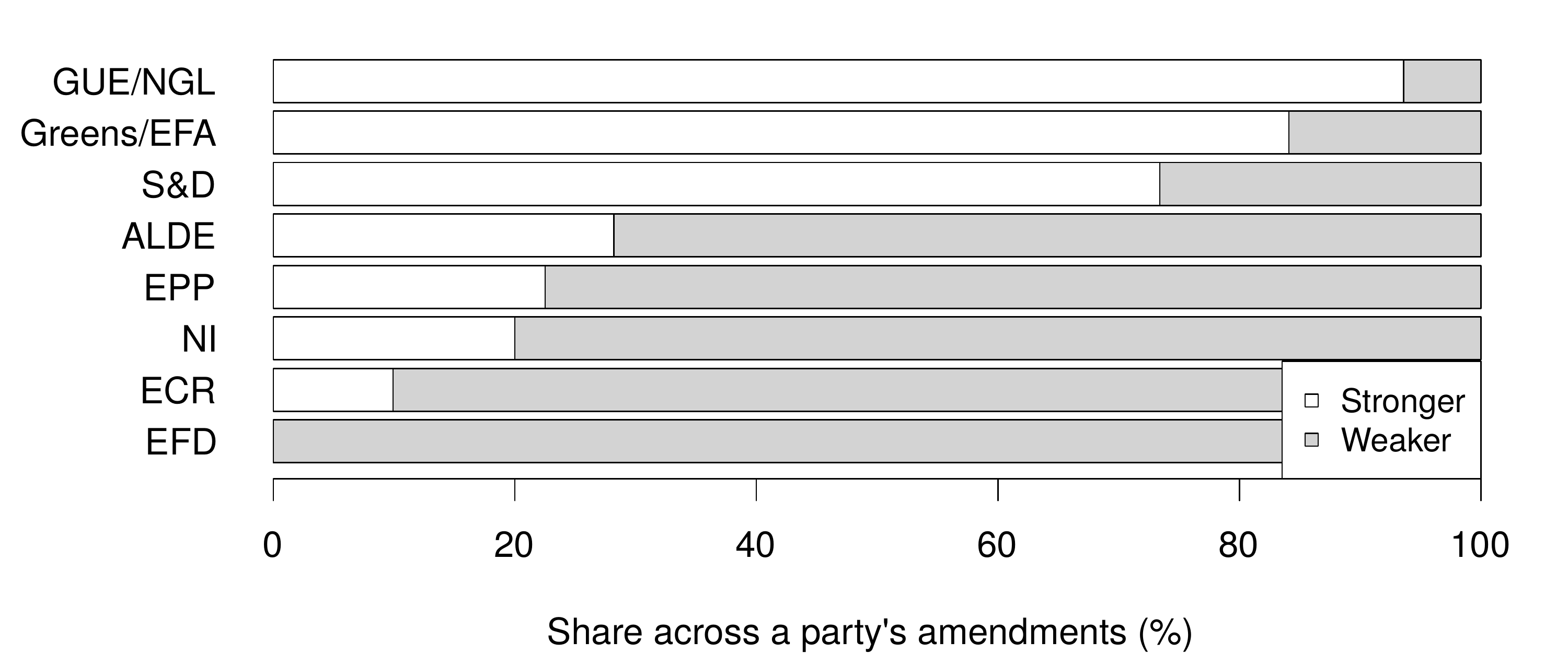}
\caption{Share of amendments ($n = 2179$) made by members of European political parties that either strengthened or weakened the GDPR's privacy and data protection requirements according to \citet{lobbyplag19a}}
\label{fig: amendments}
\end{figure}

First, the channeling tends to reflect party lines and coalitions also in privacy politics, although it remains unclear how well EU parties and their parliamentarians themselves articulate their goals for privacy policies. For instance, the legislative drafting of the GDPR tended to follow conventional party lines rather neatly. According to data from a civil society group (see Figure~\ref{fig: amendments}), the left-wing block (GUE/NGL and S\&D) together with Greens/EFA mostly made amendments that strengthened the legislation's privacy and data protection requirements, whereas the remaining parties mostly weakened these.

Second, the amending phase is only one of the stages during which lobbying occurs in the union. Before amendments are made to a draft introduced by the Commission via a rapporteur, there is a lengthy open consultation phase. This phase is typically the kernel for lobbying. During the amending phase lobbyists tend to avoid pressing legislators who are known to be unfriendly to the policy position advocated~\citep{Marshall10}, which might explain also the cross-party distribution in Figure~\ref{fig: amendments}. During the consultation phase all bets are off. The length of the phase is also important; there is plenty of time to channel interests in EU politics. For instance, the GDPR's preparations trace all the way back to the Treaty of Lisbon. Open consultations were held between 2009 and 2012, and Jan Philipp Albrecht (Greens/EFA) proposed the legislation in 2013. The privacy politics during the consultation did not follow party lines; rather, the dividing line was between industry and civil society \citep{Minkkinen19}. Both sides also actively adopted strategies for coalition building in order to maximize their power to influence the outcome.

Third, privacy politics are marked by a resource imbalance. The lobbying endeavors of a just few technology giants alone is a ``multi-million-dollar enterprise'' whose tentacles have a global reach~\citep{Popiel18}. Money is not the only important resource for power, however. Further sources include things like the geographic origin of an interest group, its internal structure and cohesion, and its access and proximity to elites, whether those in Brussels or those in Capitol~\citep{Dur08, Naoi09}. There are a couple of particularly important points embedded to the previous listing: elites and geography. 

In terms of the former, the close relationship between industry and the Commission in particular \change{has} resulted a distinct institutionalized governance form best characterized as elite pluralism~\citep{Coen98}. Powerful bedfellows have powerful vested interests. In this regard, (re)reading \citet[pp.~266--267]{Mills56} is always a worthwhile investment; the lack of independence between the two weakens checks and balances, and, in the context of the European Union, the democratic accountability of the supranational political system. In terms of geography, the vested interests obviously almost always vary across the member states, but this is not the only important element of geographic variance. It is important to have direct presence in policy hubs in order to safeguard interests~\citep{Karnikova12}. It may be also important to have a patron in some particular member states to influence the supranational politics through national lobbying. And it may be important to build distinct European coalitions even though the interests lobbied are international. In fact, none of these spatial points mean that the interests lobbied would vary much between European and multinational companies. Previous results indicate that the industry's voice is global also in EU lobbying~\citep{Bernhagen09}. The message voiced is highly similar everywhere: economy is suffering; self-regulation should be the norm; hoarding of personal data is a legitimate interest as the data is never misused; data minimization and anonymization are impossible; consumers are rational; opt-out must be the default; consent is all that is needed; regulations are expensive and inflexible; sanctions should never be used; and so forth \citep{Dix13, Fuchs17b, Minkkinen19}. Given that many of these issues are also highly technical, it is necessary to proceed to a different kind of a resource for power---knowledge is power.

Fourth, lobbying is always also about shaping the beliefs, ideas, and cognition of those lobbied \citep{Dur08, Karnikova12}. This aspect is particularly important in privacy politics. Given the limited resources to obtain knowledge, the Commission in particular has had to rely on impervious  consultation with private interests in order to ensure its own legitimacy and to acquire technical knowledge for the legitimization \citep{Broscheid03, Coen98}. Given the complex technical nature of many privacy issues, such as those related to profiling and fingerprinting, this reliance on interest groups has made it easy to feed a simple piecemeal narrative to complex socio-technical problems. Nothing to hide, and so forth. Whether it is (information) security or privacy, a reoccurring theme in the criticism expressed by the global technical community, including computer science, is indeed the extremely poor understanding of technology and the impact of the decisions made without properly understanding the consequences. This general lack of technical knowledge is also actively exploited by all parties involved in privacy politics:

\begin{quote}
``LobbyPlag.eu was started as a project to track “copy and paste” data protection legislation in the European Parliament. A number of Parliamentarians have simply copied changes to data protection laws from Amazon, eBay, the American Chamber of Commerce, but also from Privacy NGOs -- consequently called `lobby and paste'.'' \citep{lobbyplag19a}
\end{quote}

Last, an often overlooked aspect is that lobbying does not end to an enforcement of a legislation. Privacy is a good example of incremental politics that span decades, not years or months. As a good example: immediately after the GDPR was enforced, a different kind of a lobbying started via consultants, law firms, and other industry groups who managed to intentionally miscommunicate the regulation to the public by building a negative narrative around the perceived complexity and implementation difficulties~\citep{Panoptykon19a}. The ultimate goal of the Commission was to harmonize data protection regulation in the EU once and for all \citep{Dix13}, but this message was quickly lost to the public noise. While some of the agents in this post-enforcement lobbying had their own profit-seeking motives, others prepared the political landscape for things to come. In other words, it was well-known already long before the GDPR's enforcement that another legislation was in the pipeline: the bootless ePrivacy legislation from 2002 was about to be updated. Actually, the leaked draft in late 2016 revealed that ePrivacy was originally intended to be introduced together with the GDPR as a unified package~\citep{Burden17}. At the time of writing, it seems that this indirect lobbying largely driven by falsehoods and fear mongering has also been successful: the implementation of ePrivacy has stalled and its future is unclear, but the relentless lobbying nevertheless continues~\citep{EDRI19a, AccessNow19a}. For the purposes of the forthcoming empirical analysis, it is therefore important to emphasize that power resources are not resources that can be simply allocated for a task and then canceled. 

\section*{Dress Rehearsal}

\subsection{Data}

The dataset is based on the so-called Transparency Register (TR) maintained by the \citet{EU19a} for keeping track of lobbyists. After a number of early 1990s initiatives and registers for improving the transparency of lobbying~\citep{Greenwood13}, the TR was launched in 2011 as a joint venture of the Parliament and the Commission. Although disclosure of information to the TR is voluntary, in practice, it represents most companies, associations, and other organizations, as well as individuals, who have an interest to influence policy making in the EU. It is also the largest register of its kind; presently, it covers over 12 thousand active registrants~\citep{EU19b}. Even though a snapshot taken in June 2019 is used, some observations refer also to the years 2018 and 2017 due to update delays, accounting practicalities, and related issues.

The dataset contains only those who have declared having an interest in privacy and data protection issues. In practice, the required subsetting was carried with a simple keyword search from the fields denoting the self-declared goals of a registrant and the predefined activities covered in the TR. Initially, three keywords were used for the case insensitive searches: \textit{data protection}, \textit{gdpr}, and \textit{privacy}. However, many registrants have supplied their voluntary disclosures in their native languages. To account for these cases, the three keywords were further machine-translated to all of the 24 official languages in the EU by using Google's online translation service. Each translated keyword was subsequently searched from the two fields. While this automated solution is hardly perfect, it seems sufficient for the simple task of enlarging the amount of observations. After all, the most important lobbyists supply their voluntary disclosures in English.

\begin{table}[th!b]
\centering
\caption{Categorization}
\label{tab: groups}
\begin{tabular}{ll}
\toprule
Group & Included subcategories (original TR-labels) \\
\hline
Companies and business & $\bullet$ Companies \& groups \\
associations $(n = 616)$ & $\bullet$ Trade and business associations \\
\cmidrule{2-2}
Employee associations $(n = 70)$ & $\bullet$ Trade unions and professional associations \\
\cmidrule{2-2}
Law firms and & $\bullet$ Law firms \\
consultants $(n = 115)$ & $\bullet$ Self-employed consultants \\
& $\bullet$ Professional consultancies \\
\cmidrule{2-2}
Non-governmental & $\bullet$ Non-governmental organisations, \\
organizations $(n = 166)$ & platforms and networks and similar \\
\cmidrule{2-2}
Public authorities $(n = 29)$ & $\bullet$ Regional structures \\
& $\bullet$ Other sub-national public authorities \\
& $\bullet$ Other public or mixed entities, created by law \\
& whose purpose is to act in the public interest \\
& $\bullet$ Transnational associations and networks of public \\ 
& regional or other sub-national authorities \\
\cmidrule{2-2}
Research institutions $(n = 44)$ & $\bullet$ Academic institutions \\
& $\bullet$ Think tanks and research institutions \\
\bottomrule
\end{tabular}
\end{table}

The empirical analysis is based on the sub-categories provided in the TF for classifying the registrants. After excluding five organizations representing churches and religious communities as well as 27 unclassified organizations, the sub-categories were merged into the six groups enumerated in Table~\ref{tab: groups}. This regrouping is contextually sensible. It also ensures a sufficient amount of observations in each group. Furthermore, these six groups in the assembled privacy-specific dataset reflect the overall distribution in the TR database~\citep{EU19b}. The largest group refers to companies and business associations (CBAs). Non-governmental organizations (NGOs), which are taken as representatives of the European civil society, is the second last group in the dataset.

\subsection{Methods}

Four power resources are considered. The first two are classical \citep{Korpi85}, the third one is specific to the European Union, and the fourth relates to implicit coalitions and networking. The four resources and their operationalization can be elaborated as follows:

\begin{enumerate}
\item{\textit{Money} talks also in lobbying. In addition to testifying the general importance of money for representing interests, previous results have shown that money affects the probability of having an office in Brussels and the amount of information supplied to the Commission, among other things~\citep{vanHecke18}. To measure lobbying budgets, the self-disclosed annual monetary amounts for the activities covered in the TR are used. Two details are worth further mentioning:  those cases are excluded that use a currency other than euro for reporting the annual costs, and the maximum values are used in case the costs are reported as a range.}
\item{\textit{Human resources} are also important. Lobbying is a labor-intensive occupation. Thus, the total number of persons involved in the activities covered in the TR are used. Although the TR provides fields for reporting also the share of working time allocated to lobbying activities, not all registrants have disclosed this information consistently. The number of persons involved is therefore a more robust measure.}
\item{\textit{Geography} affects lobbying in the EU. It also correlates with financial and human resources; establishing a permanent office in Brussels is expensive and requires hiring EU specialists~\citep{DeBruycker18}. Following existing research~\citep{Hollman18}, the contact details provided are used to record the registrants' countries of origin; the records do not refer to contact details of a potential Brussels office. Thus, Facebook, for instance, is located to Ireland.}
\item{\textit{Networking} with others is important in order to derive new power resources (including funding) through explicit or implicit coalitions stemming from memberships in further associations. 
Network-based power does not depend only on networking with others of similar kind. Therefore: for each registrant in the privacy-specific subset, an undirected and unweighted network was constructed by connecting the registrant to all associations, organizations, or other groups to which the given registrant disclosed belonging to, regardless whether these are privacy-specific or not. Alas, the TR does not provide a structured format for the memberships. For this reason, the construction was limited to those specific associations, organizations, or other groups that were also registered to the TR. For each registrant, the matching was done with case-sensitive searches for the names of all other registrants from a free-form membership field of the given registrant.}
\end{enumerate}

The analysis is carried out primarily with descriptive statistics. To compare group averages, the non-parametric test of \citet{KruskalWallis52} is used. In addition, the analysis of variance (ANOVA) method of \citet{ScottKnott74} is used for briefly examining further clustering of the groups according to group means. Despite of some limitations, the method has been argued to perform well in comparison to other multiple comparison procedures~\citep{Idri16}. Standard techniques are used also for the network computations. To compare individual nodes (registrants) and the groups they belong to, node degrees and so-called node betweenness are used. Both are classical measures~\citep{Freeman77}. The former gives the number of other registrants to which a registrant is connected through self-disclosed information about memberships and affiliations. The latter is roughly defined by the number of shortest paths going through a node; it is a typical measure for probing connectivity and network hubs. Finally, cross-group connections are examined by computing the fraction of a group's connections connected to other groups scaled by the group's all connections. Although the labels and technical details vary slightly \citep{EverettBorgatti05, Sluban18a}, also this group-based computation belongs to a standard toolbox for applied network analysis.

\section*{Limelight}

\subsection{Financial and Human Resources}

\begin{figure}[th!b]
\centering
\includegraphics[width=\linewidth, height=5cm]{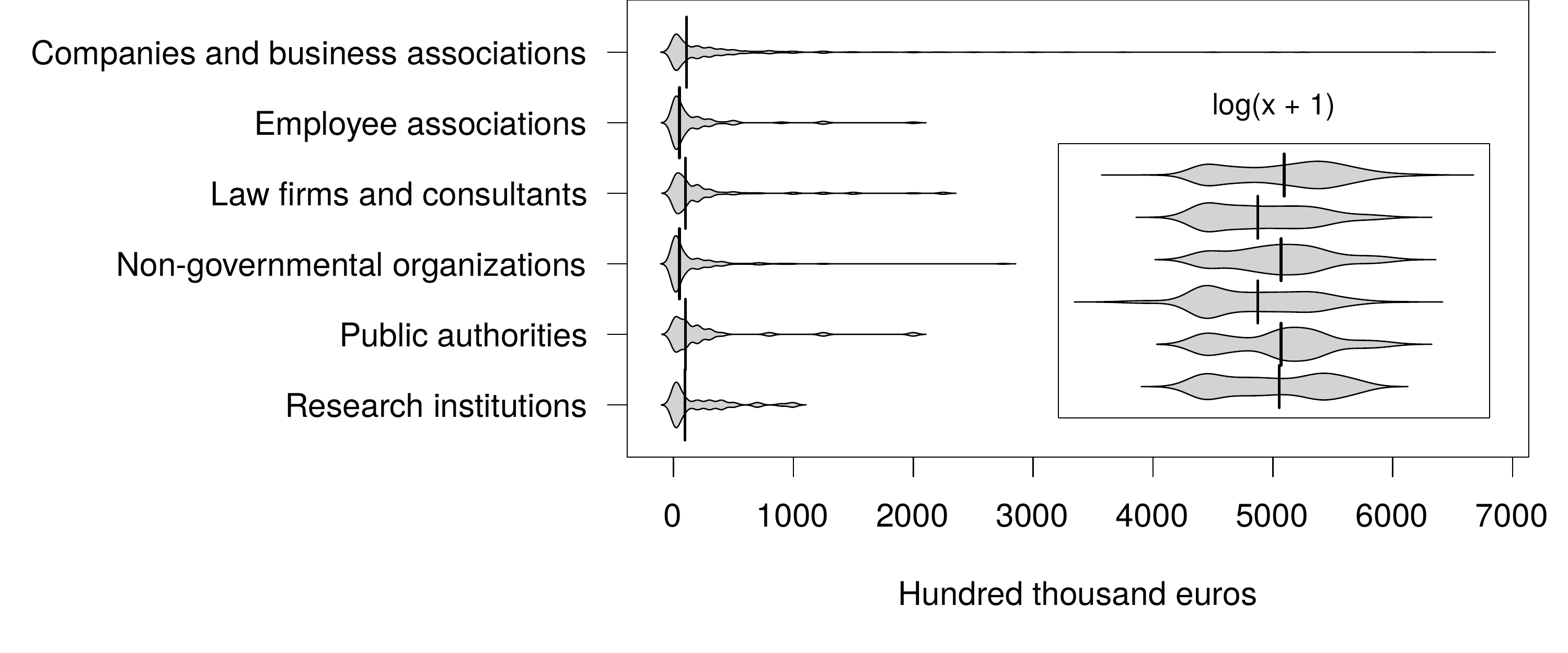}
\caption{Lobbying budget across groups}
\label{fig: bean budget}
\end{figure}

\begin{figure}[th!b]
\centering
\includegraphics[width=\linewidth, height=5cm]{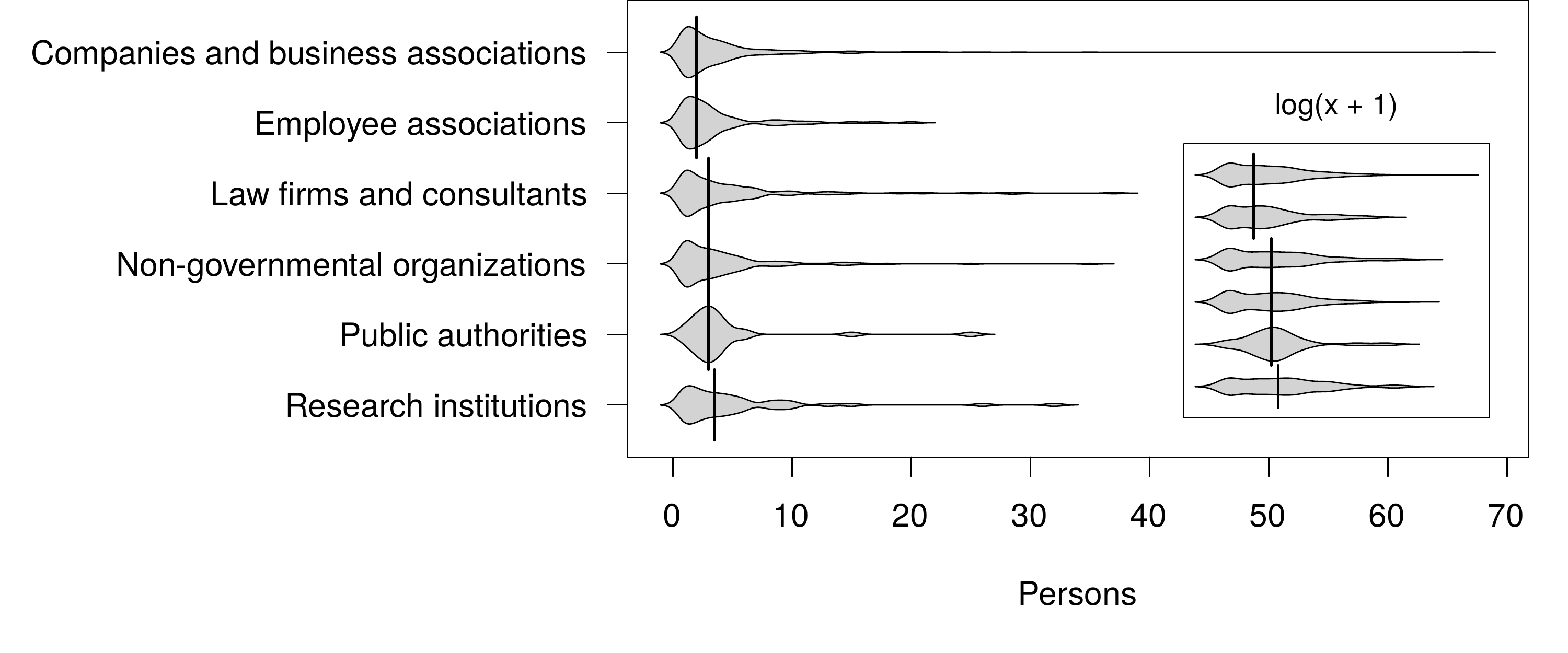}
\caption{Lobbying personnel across groups}
\label{fig: bean persons}
\end{figure}

Financial and human resources are also basic power resources for efficient lobbying. Thus, Figures~\ref{fig: bean budget} and \ref{fig: bean persons} visualize the lobbying budgets and human resources across the six groups with so-called ``beanplots'' \citep{beanplot}. Each bean for a group illustrates the distribution of observations within the group. Vertical lines denote group medians. As can be seen, the shape of the distributions is rather similar in all beans. Due to the lengthy tails of the  distributions, the similarity is better seen from the inner plots that visualize the same information with a logarithm transformation. Also the medians are quite close to each other, although formal evaluation with the Kruskal-Wallis test rejects the null hypothesis of equal medians for the lobbying budgets ($\chi^2(5) \simeq 37.8; p < 0.001$) but not for the personnel ($\chi^2(5) \simeq 7.4; p \simeq 0.195$). This rejection is attributable to the NGOs and employee associations who both have slightly smaller budgets on average.

\begin{table}[th!b]
\centering
\caption{Clusters between groups (Scott-Knott)}
\label{tab: scott-knott}
\begin{tabular}{lccrcc}
\toprule
Group & \multicolumn{2}{c}{Budget} && \multicolumn{2}{c}{Personnel} \\
\cmidrule{2-3}\cmidrule{5-6}
& $x$ & $\log(x + 1)$ && $x$ & $\log(x + 1)$ \\
\hline
Companies and business associations & A & A && A & A \\
Employee associations & B & B && A & B \\
Law firms and consultants & B & B  && B & B \\
Non-governmental organizations & B & B && B & C \\
Public authorities & C & C && B & C \\
Research institutions & C & D && B & C \\
\hline
Number of clusters & 3 & 4 && 2 & 3 \\
\bottomrule
\end{tabular}
\end{table}

Another way to look at the six groups is to consider the clustering between them. The Scott-Knott multiple comparison test is suitable for this task. The results from the test are summarized in Table~\ref{tab: scott-knott}. As the test requires the conventional normality assumption due to the reliance on ANOVA, the computation was done also with the same logarithm transformation used for the beanplots. The clusters are relatively clear for the companies and business associations observed; these form their own cluster in all cases except lobbying personnel when the logarithm transformation is not applied. In terms of financial resources, NGOs cluster together with employee associations, law firms, and consultants. While the results are less clear with respect to human resources, the main point from the Scott-Knott computations is the distinctiveness of CBAs and their resources. It is possible to consider these resources also from a different angle.

\begin{table}[p!]
\centering
\caption{Top-25 Companies and Business Associations According to Lobbying Budget}
\label{tab: top25companies}
\begin{tabularx}{\linewidth}{Xll}
\toprule
Company or association & Budget & Personnel \\
& (1000~\euro) & (persons) \\
\hline
Insurance Europe & 6749 &  35 \\ 
Google $\bullet$ & 6250 &  15 \\ 
Microsoft Corporation $\bullet$ & 5250 &  15 \\ 
Association for Financial Markets in Europe & 5000 &  67 \\ 
European Banking Federation & 4500 &  22 \\ 
Facebook Ireland Limited $\bullet$ & 3749 &  20 \\ 
Deutsche Bank AG & 3287 &   9 \\ 
Association des Constructeurs Europ\'eens d'Automobiles & 3000 &  19 \\ 
IDIADA Automotive Technology, S.A & 3000 &  15 \\ 
Gesamtverband der Deutschen Versicherungswirtschaft e.V. & 2749 &  21 \\
Daimler Aktiengesellschaft & 2500 &  16 \\                                                                                    
Novartis International AG & 2500 &  15 \\                                                                                     
Apple Inc. $\bullet$ & 2250 &   7 \\                                                                                                   
Huawei Technologies & 2190 &  10 \\                                                                                          
General Electric Company & 2000 &   8 \\                                                                                      
IBM Corporation & 2000 &   9 \\                                                                                               
Vodafone Belgium SA & 2000 &   8 \\                                                                                           
Amazon Europe Core SARL $\bullet$ & 2000 &  10 \\                                                                                      
UK Finance Limited & 2000 &  15 \\                                                                                          
Telefonica, S.A. & 1800 &   6 \\                                                                                                
GlaxoSmithKline & 1749 &  11 \\                                                                                               
Deutsche Telekom & 1690 &   9 \\                                                                                              
Association of British Insurers & 1500 &   4 \\ 
Qualcomm Inc. & 1500 &   6 \\ 
Bank of America Merrill Lynch & 1500 &   5 \\ 
\bottomrule
$\bullet$ GAFAM
\end{tabularx}
%
\vspace{15pt}
%
\centering
\caption{Top-10 Non-Governmental Organizations According to Lobbying Budget}
\label{tab: top10ngos}
\begin{tabularx}{\linewidth}{Xll}
\toprule
Organization or association & Budget & Personnel \\
& (1000~\euro) & (persons) \\
\hline
Bureau Europ\'en des Unions de Consommateurs & 2749 &  35 \\ 
First Draft & 1250 &   3 \\ 
Civil Liberties Union for Europe & 980 &  10 \\ 
European Organisation for Rare Diseases & 900 &  14 \\ 
European Patients' Forum & 767 &   9 \\ 
European co-operation for Accreditation & 723 &   1 \\ 
Friedrich-Naumann-Stiftung f\"ur die Freiheit & 700 &   6 \\ 
The Mentor Group & 700 &   8 \\ 
The NHS Confederation & 600 &   4 \\ 
Center for Democracy \& Technology & 500 &   4 \\ 
\bottomrule
\end{tabularx}
\end{table}

\begin{figure}[th!b]
\centering
\includegraphics[width=\linewidth, height=4cm]{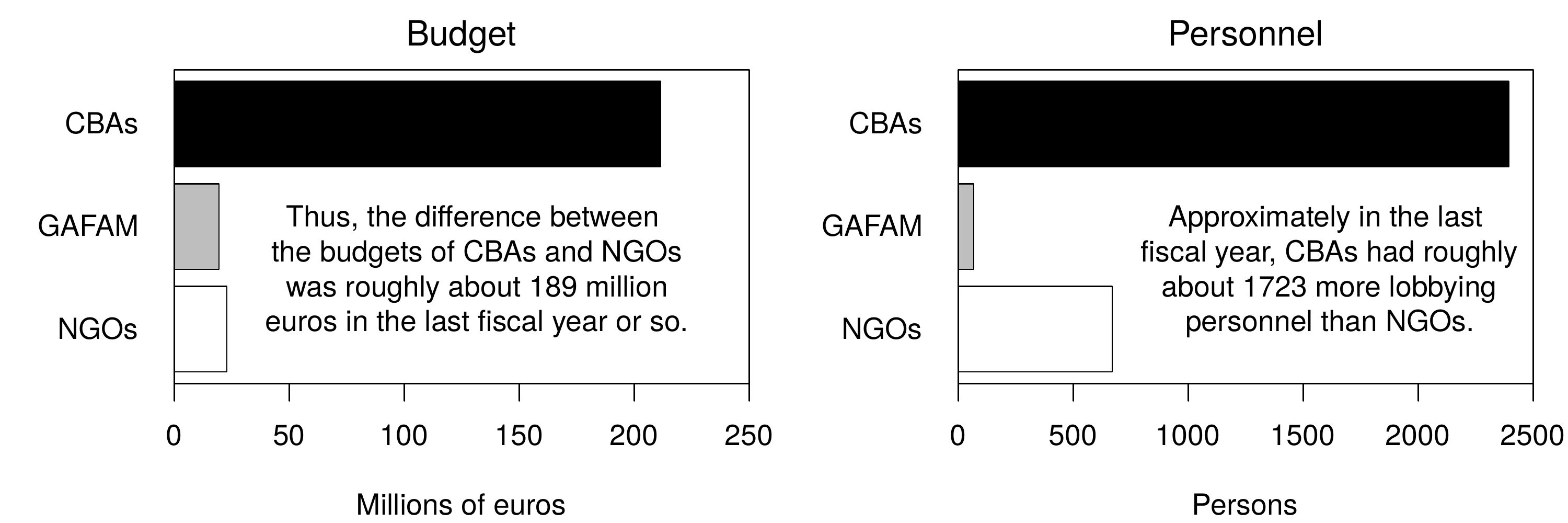}
\caption{Absolute amount of lobbying resources of three groups}
\label{fig: absolute}
\end{figure}

It is debatable whether mean, median, or any other measure for the central tendency makes sense in measuring power resources. It is equally sensible to argue for the use of absolute values: if members of a group activate their power resources for the use of power, and a perfect consensus is assumed to exist within the group, the absolute amounts of financial and human resources would arguably be a much better proxy than the group's resources on average. Power resources are important also without being activated \citep{Korpi85}. To indirectly influence the Commission, the Parliament, or even the Council (which, however, is not participating in the TR), it may be sufficient in some cases to merely demonstrate the possession of vast resources. Given this reasoning, Figure~\ref{fig: absolute} shows the absolute amounts for three groups. The middle group refers to the so-called GAFAM (Google, Amazon, Facebook, Apple, and Microsoft), which currently dominates the global Internet market capitalization~\citep{Paul18}. The differences are substantial between the CBAs and NGOs. In terms of budgets allocated for lobbying, furthermore, the GAFAM group is only slightly behind the total combined budget of all 166 non-governmental organizations observed. By further adding IBM, say, to the GAFAM(I) group, its budget is already bigger than that of all NGOs in the sample.

The CBAs with the largest lobbying budgets enumerated in Table~\ref{tab: top25companies} are also interesting. While the GAFAM group is expectedly present, the listing misses some notable business players in privacy politics, such as the Interactive Advertising Bureau (IAB) and its European branches; IAB is also at the epicenter of many GDPR controversies~\citep{Brave19a}. In fact, IAB's main European branch in Brussels ``only'' has a lobbying budget of about 399 thousand euros to be used for the association's five mostly full-time lobbyists. When combined with the branches in Poland and Slovakia, the combined amount is much more than what the majority of civil society groups and their coalitions have, sans the few exceptions ranked in Table~\ref{tab: top10ngos}. This ranking is likewise interesting. General rights for consumers and journalists, as well as human rights, are well-represented, holding the top-three positions in terms of financial resources. However, NGOs specialized to privacy and data protection do not appear in the list. Only upon examining a top-50 ranking do privacy-specific groups, such as European Digital Rights (EDRi) and Access Now Europe, show in a listing. This observation serves to emphasize an important point: the NGOs with interests in privacy politics constitute a highly heterogeneous group. Having a unified stance on privacy matters may be difficult; EDRi's interests may not necessarily correlate well with interests of NGOs specialized to healthcare, science, social security, sports, or even tourism. This lack of cohesion may also weaken coordinated use of the scarce resources. In any case: with respect to organized business interests and their vast power resources for lobbying, the traditionally generous funding provided by the EU for various different non-governmental organizations~\citep{Greenwood13, Ram11} hardly seems sufficient for balancing the resources for power in privacy politics.

\subsection{Geography}

\begin{figure}[th!b]
\centering
\includegraphics[width=\linewidth, height=12cm]{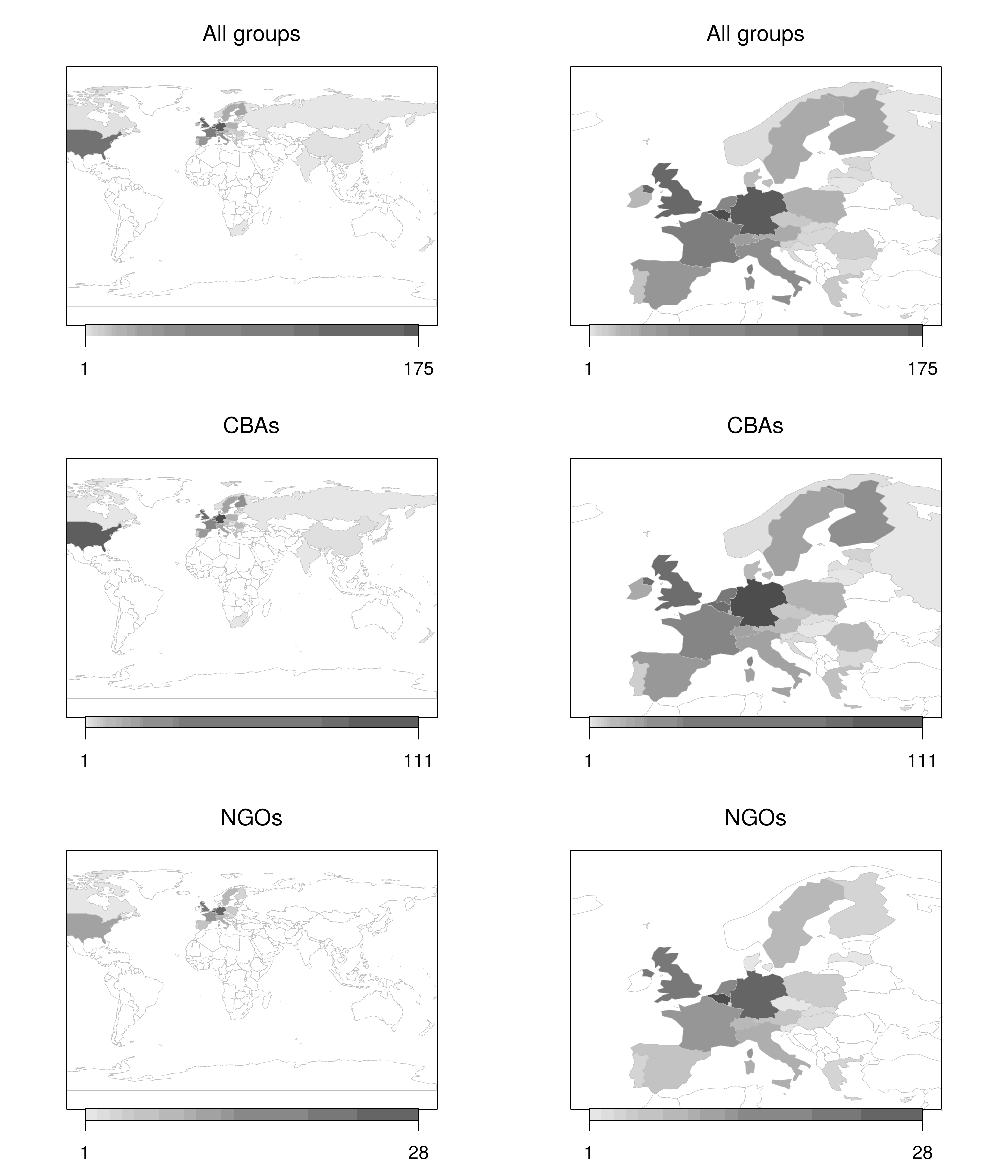}
\caption{Geographic origin of the groups}
\label{fig: maps}
\end{figure}

The absolute amounts of financial and human resources for lobbying vary substantially between organized business and civil society interests. However, there may be a certain balancing effect because many of the NGOs involved in European privacy politics are also from Europe. To illustrate this observation, Figure~\ref{fig: maps} shows the disclosed geographic origins of all groups in the sample, as well as the locations of CBAs and NGOs.

The two plots on the first row demonstrate the sample's general and expected bias toward Europe. However, the United States ranks fourth $(n = 100)$, right after the United Kingdom $(n = 123)$, Germany $(n = 167)$, and Belgium $(n = 175)$.  Interestingly, the subsequent plots on the second row foretell that it is particularly companies and business associations who are registered to the TR but often still located in the United States. From all CBAs in the sample, about 14\% have origins in the United States; only Germany ranks higher (18\%). When the focus is shifted to the NGOs and the two plots on the third row, it is clear that an analogous effect is much less pronounced; most of the NGOs observed are European. Germany and Brussels (Belgium) are particularly well-represented in this regard. Although the link remains implicit with a focus on power resources alone, this observation is expected: much of the privacy and data protection legislation in the EU has been modeled according to the national legislation in Germany, and, therefore, it is to be expected that also the voice of the German civil society is loud in privacy politics. That said, it should be emphasized that also more fine-grained regional aspects affect lobbying within the EU~\citep{vanHecke18}. As many legislators specialize in regional matters, so do lobbyists. Operating at the EU-level requires aggregating heterogeneous interests~\citep{DeBruycker18}, which may also vary geographically. However, it remains generally unclear whether privacy politics vary within the EU. In terms of organized business interests, the variance is likely small, but things may be different with various national, regional, and local public authorities, research institutions, and even European NGOs due to their heterogeneous composition. In any case, these geographic and regional viewpoints to privacy politics provide a good question for further work.

\subsection{Networks}

\begin{figure}[th!b]
\centering
\includegraphics[width=\linewidth, height=4cm]{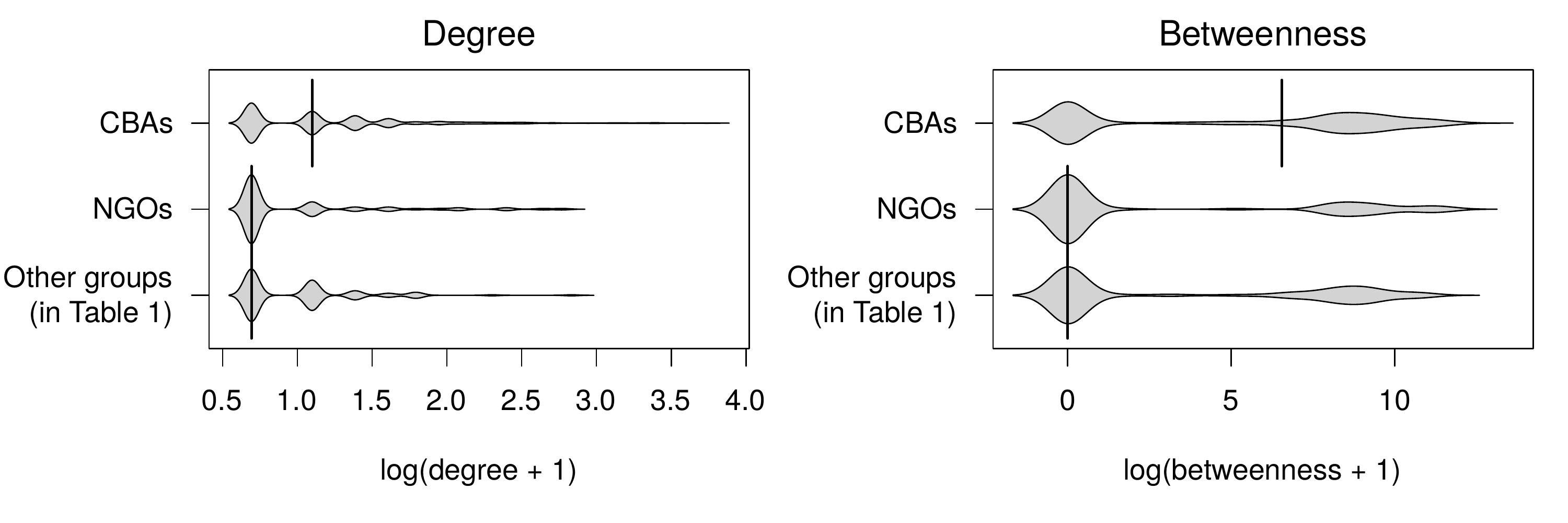}
\caption{Network centrality across three groups}
\label{fig: centrality}
\end{figure}

There exists variation between financial resources, human resources, and geography. To examine whether the same holds further in terms of memberships and affiliations, the two node centrality measures are sufficient. The results are shown in Figure~\ref{fig: centrality}. Although not made explicit, it should be recalled that the network constructed contains also lobbyists who do not have an interest in privacy and data protection. While keeping this point in mind, the left-hand side plot indicates that many registrants in the six groups observed have just one affiliation. However, the median is higher for the CBAs. The same holds for the node betweenness scores shown on the right-hand side plot. Also the two hypotheses about equal medians are expectedly rejected according to Kruskal-Wallis tests (${\chi^2(2) \simeq 20.3, \chi^2(2) \simeq 15.2, p < 0.001}$). Companies and business associations tend to excel also at networking better than NGOs and other groups.

\begin{figure}[th!b]
\centering
\includegraphics[width=8cm, height=8cm]{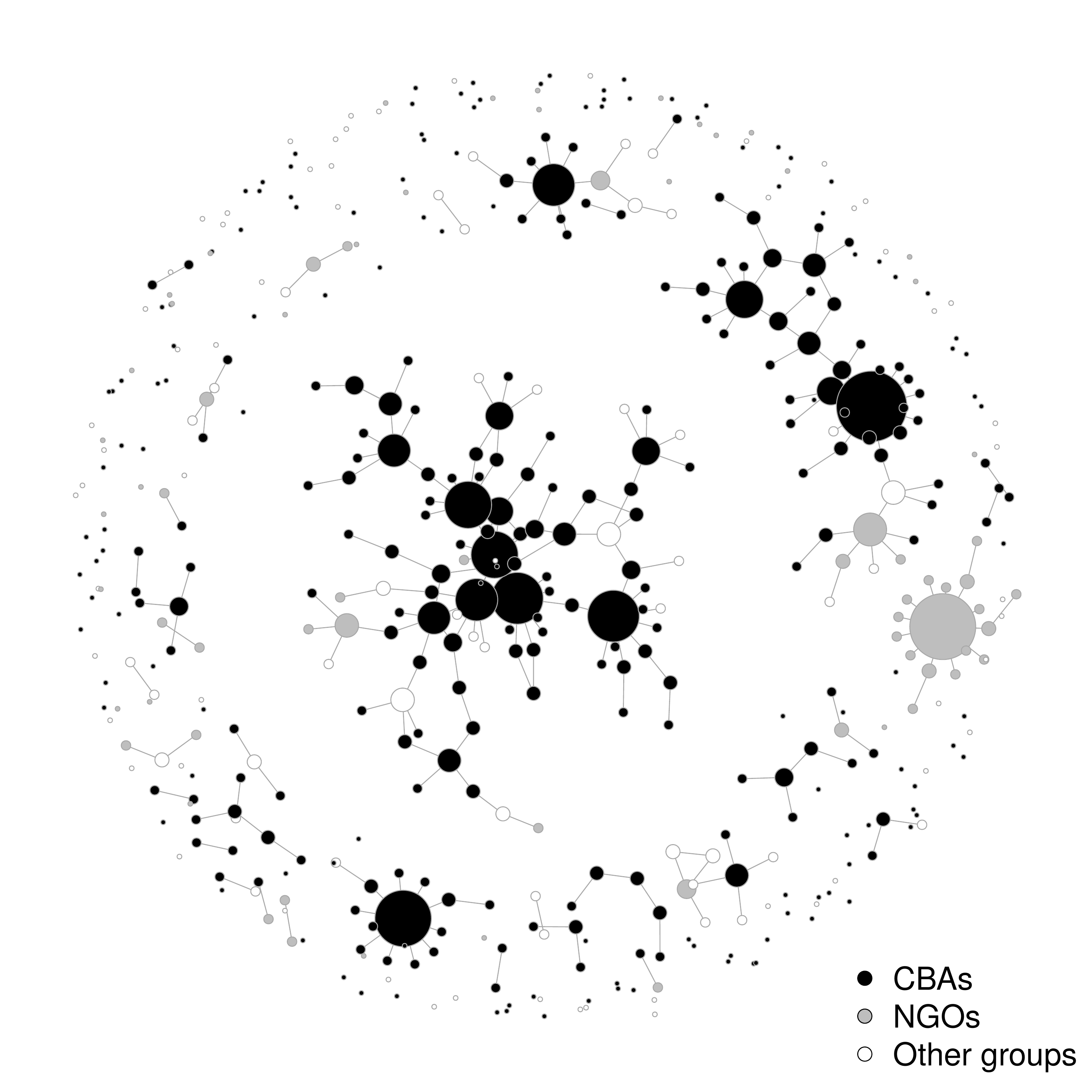}
\caption{Induced network with nodes scaled by their degrees}
\label{fig: network}
\end{figure}

\begin{table}[p!]
\centering
\caption{Top-17 Companies and Business Associations According to Network Degree}
\label{tab: top17companiesnetwork}
\begin{tabularx}{\linewidth}{Xc}
\toprule
Organization or association & Degree \\
\hline
European Banking Federation & 41 \\                                                                                                                         
Insurance Europe & 35 \\                                                                                                                                  
World Federation of Advertisers & 29 \\                                                                                                                   
International Chamber of Commerce & 29 \\                                                                                                                 
Association for Financial Markets in Europe & 26 \\                                                                                                       
American Chamber of Commerce to the European Union & 23 \\                                                                                                
MedTech Europe & 21 \\                                                                                                                                    
DIGITALEUROPE & 18 \\                                                                                                                                     
Creativity Works! & 14 \\                                                                                                                                 
ABB & 14 \\                                                                                                                                               
European Advertising Standards Alliance & 12 \\                                                                                                           
European Publishers Council & 12 \\                                                                                                                       
European Telecommunications Network Operators' Association & 11 \\                                                                                        
Microsoft Corporation $\bullet$ & 11 \\                                                                                                                             
Institute of International Finance & 11 \\                                                                                                                
Google $\bullet$ & 11 \\                                                                                                                                            
Leaseurope & 10 \\   
\bottomrule
$\bullet$ GAFAM
\end{tabularx}
%
\vspace{15pt}
%
\centering
\caption{Top-17 Non-Governmental Organizations According to Network Degree}
\label{tab: top17ngosnetwork}
\begin{tabularx}{\linewidth}{Xc}
\toprule
Organization or association & Degree \\
\hline
European Digital Rights & 15 \\ 
European Patients’ Forum & 13 \\ 
Bundesverband IT-Sicherheit e.V. (TeleTrusT) & 10 \\ 
International Federation of Reproduction Rights Organisations & 10 \\ 
European Organisation for Research and Treatment of Cancer & 7 \\ 
International Diabetes Federation European Region & 7 \\ 
Alliance for Lobbying Transparency and Ethics Regulation & 6 \\ 
April & 5 \\ 
Corporate Europe Observatory & 4 \\ 
Global Network Initiative & 4 \\ 
International Confederation of Societies of Authors and Composers & 4 \\ 
Bureau Europ\'een des Unions de Consommateurs & 3 \\ 
ICON & 3 \\ 
Helsinki Foundation for Human Rights & 3 \\ 
Open Rights Group & 2 \\ 
Arbeitsgemeinschaft f\"ur wirtschaftliche Verwaltung e.V. & 2 \\ 
Privacy International & 2 \\ 
\bottomrule
\end{tabularx}
\end{table}

As with the financial and human resources, both centrality measures exhibit also a very long tail, indicating the presence of a few particularly well-connected network hubs. To illustrate these hubs visually, Figure~\ref{fig: network} shows a reduced network using the algorithm of \citet{KamadaKawai89}. The reduction refers to the omission of all nodes not present in the privacy-specific subset analyzed in the previous sections; therefore, the disconnected nodes shown mostly refer to registrants who only have affiliations with organizations, associations, or other groups not having declared interests in privacy and data protection. While again keeping these small technicalities in mind, the presence of powerful hubs is indeed highly vivid in the figure. None of these are particularly well-connected to other powerful hubs. Most of these are CBAs. In fact, there is only one large NGO hub located in the southeast corner of the visualized network. It maps to EDRi, which is connected to many other privacy NGOs, including such examples as Open Rights Group, Access Now Europe, Electronic Frontier Finland -- Effi ry, Fundacja Panoptykon, Chaos Computer Club e.V., Digitalcourage e.V., and Privacy International. 

In other words, EDRi in particular faces much better at networking than in terms of financial resources. The degree-based rankings in Tables~\ref{tab: top17companiesnetwork} and \ref{tab: top17ngosnetwork} indicate a more general divergence between those who possess particularly large lobbying budgets and those who are particularly well-connected. While the big European associations representing the banking and financial sectors are represented in both lists, many digital advertisement associations only appear in the degree-based ranking for CBAs. Another point is that only Google and Microsoft are represented from the GAFAM group. The explanation may be simple; the companies in the group are so powerful that networking may not be a necessity for deriving further resources to exercise power. An analogous explanation may apply on the NGO-side; the lobbying budgets of the consumer, journalism, and human rights groups in Table~\ref{tab: top10ngos} may be sufficient to avoid wasting effort to network with other NGOs. Furthermore, the American Chamber of Commerce to the European Union appears in the network ranking for CBAs. Given the earlier comments about the lobbying practices witnessed with the GDPR's parliamentary amending, it thus seems that good connections may be good also for the ``lobby and paste'' strategy. All this said, the bottom line is that CBAs are also better at networking---or have more resources to do so---compared to non-governmental organizations. While the results are partially explained by the fact that there are also more CBAs in the sample (see Table~\ref{tab: groups}), it seems reasonable to assume that the principles and practices for networking are more mature and well-established for companies and business associations. This assumption is generally backed by the enormous amount of historical evidence presented by the comparative VoC school.

\begin{table}[th!b]
\centering
\caption{Cross-Group Network Connections}
\label{tab: cross-group}
\begin{tabular}{lrrrr|r}
\toprule
& Non-privacy & CBAs & NGOs & Other groups & $\sum$ \\
\hline
Non-privacy & 7950 & 694 & 90 & 150 & 8884 \\ 
CBAs & 694 & 330 & 13 & 34 & 1071 \\ 
NGOs & 90 & 13 & 44 & 14 & 161 \\ 
Others (in Table~\ref{tab: groups}) & 150 & 34 & 14 & 12 & 210 \\ 
\hline
$\sum$ & 8884 & 1071 & 161 & 210 & 10326 \\ 
\bottomrule
\end{tabular}
\end{table}

The final result presented relates to so-called unholy alliances. Previous results indicate that particularly controversial political questions that attain sweeping media attention tend to increase the probability of strange lobbying bedfellows \citep{BeyersDeBruycker18}. While privacy is a prime example about politically sensitive issues, already the visualization in Figure~\ref{fig: network} hints that beds are seldom shared with antagonists at least in terms of official cross-group affiliations. Table \ref{tab: cross-group} makes the observation explicit. About 31\% of all connections of the CBAs observed trace to other CBAs and only about one percent to NGOs. The analogous numbers are 27\% and 8\% for the NGOs observed. Last but not least, both the CBAs and NGOs observed tend to affiliate heavily with groups whose interests are not privacy-specific. This observation serves to emphasize that privacy is only one thing among the heterogeneous interests represented by the interest groups observed. Data protection enlarges the scope and digital rights \change{enlarge} it further, but these topics are still interlinked to many broader issues in the European Union.

\section*{Curtain Call}

This paper examined the question of how much more power resources do industry representatives have in comparison to civil society representatives in the context of privacy lobbying in the EU. The answer is clear: there is a great imbalance between the lobbying resources of the two sides. On average, financial and human resources do not vary much between the CBAs and NGOs observed, but a substantial difference exists when absolute amounts are observed. The resources of a few technology giants alone surpass the combined amount of resources possessed by all NGOs observed. Companies and business associations seem also better at networking with other organizations, associations, and groups. On the civil society side, privacy issues are mainly represented by a few specialized NGOs, which do not network much with other NGOs at least in terms of official affiliations. Of the basic power resources examined, only geography seems to favor NGOs: these are mostly European, whereas many CBAs keep their head offices in the United States even when lobbying privacy issues in the European Union.

All research contains limitations, of course, and this paper is no exception. To begin with the smaller problems, some data quality issues should be acknowledged. Despite of the transparency improvements that the TR has brought, the quality of data stored to the register remains a question mark; the voluntary nature of the register is an obvious issue, but there are also many smaller technical problems~\citep{Greenwood13}. Examples include unclear updating \change{procedures}, maintenance, and the lack of a structured format for some information. The last point is directly relevant to the presented paper; the network-based analysis misses many nodes because only TR-specific affiliations were used. That said, the TR's problems should not be exaggerated; there is also a strong incentive for the registrants to keep their records accurate~\citep{Hollman18}. Turning to the bigger issues, the paper's essential limitation is also its greatest strength. In other words, power resources do not tell anything about how power is used. Nor do these allow to infer about how the exercise of power affects policy outcomes. This classical problem in political science is particularly pronounced in lobbying research~\citep{Dur08}. But what the power resources approach enables is the relatively objective accounting of the basic capabilities to exercise power. Merely observing these resources in privacy politics is also enough to counter \citeauthor{Clarke19}'s \citeyearpar{Clarke19} argument  that conventional empirical research would be useless in the 21st century privacy and surveillance settings. But how to combine power resources with actual politics and the resulting policy outcomes? 

For seeking an answer to this question, a good starting point would be to address the essential limitation of the presented work in conjunction with the basic limitation of the work of \citet{Minkkinen19}, who draws from open consultations, which are closer to actual politics, but excludes power resources. There are good examples about the benefits from combining data from the TR with data on open consultations~\citep{Sluban18a}. However, it remains debatable whether even such a combination is sufficient for analyzing the whole legislative process in the European Union. In general, full-length tracing requires triangulating quantitative data with qualitative material. To do so, a good option would be to continue the well-established tradition of using expert interviews in lobbying research~\citep{BeyersDeBruycker18, Massaro19}. It would be also interesting to know how the European privacy NGOs interpret the power resource imbalance and its hypothetical relation to the use of power for influencing legislators. 

Finally, it is necessary to say something about the grand issues put forward by \citet{Zuboff19}, even if only tentatively. The high-level VoC approach may be useful for tackling also these issues. As was argued, classical theoretical concepts such as pluralism apply well also to privacy and related politics. The VoC approach has often characterized pluralism as being a policy making style that is typical to winner-take-all political systems \citep[p.~137]{MartinSwank12}. The global history of privacy, data protection, and digital rights certainly fits into a winner-\change{take}-all scheme; self-regulative \textit{laissez faire} has been the norm, and the winner has been the global technology industry and its organized interests. Yet, the GDPR signals that things are slowly changing. Such changes make it plausible to use the framing between uncoordinated and coordinated capitalism.

\balance
\bibliographystyle{SageH}

\end{document}